\documentstyle[12pt,epsfig]{article}

\newlength{\dinwidth}
\newlength{\dinmargin}
\begin{document}
\def\bold#1{\setbox0=\hbox{$#1$}%
     \kern-.025em\copy0\kern-\wd0
     \kern.05em\%\baselineskip=18ptemptcopy0\kern-\wd0
     \kern-.025em\raise.0433em\box0 }
\def\slash#1{\setbox0=\hbox{$#1$}#1\hskip-\wd0\dimen0=5pt\advance
         to\wd0{\hss\sl/\/\hss}}
\newcommand{\be}{\begin{equation}}
\newcommand{\ee}{\end{equation}}
\newcommand{\bea}{\begin{eqnarray}}
\newcommand{\eea}{\end{eqnarray}}
\newcommand{\nn}{\nonumber}
\newcommand{\dd}{\displaystyle}
\newcommand{\bra}[1]{\left\langle #1 \right|}
\newcommand{\ket}[1]{\left| #1 \right\rangle}
\newcommand{\spur}[1]{\not\! #1 \,}
\thispagestyle{empty}
\vspace*{1cm}
\rightline{BARI-TH/03-462}
\vspace*{2cm}
\begin{center}
  \begin{LARGE}
  \begin{bf}
Understanding $D_{sJ}(2317)$
 \vspace*{0.5cm}
  \end{bf}
  \end{LARGE}
\end{center}
\vspace*{8mm}
\begin{center}
\begin{large}
P. Colangelo and  F. De Fazio
\end{large}
\end{center}
\begin{center}
\begin{it}
Istituto Nazionale di Fisica Nucleare, Sezione di Bari, Italy
\end{it}
\end{center}
\begin{quotation}
\vspace*{1.5cm}
\begin{center}
  \begin{bf}
  Abstract\\
  \end{bf}
  \end{center}
\noindent We analyze the hadronic and radiative decay modes of
the recently observed $D_{sJ}(2317)$ meson, in the hypothesis that
it can be identifield with the scalar $s_\ell^P={1\over 2}^+$
state of $c \bar s$ spectrum ($D_{s0}$).
The method is based on heavy quark symmetries and
Vector Meson Dominance ansatz.
We find that the hadronic
isospin violating mode $D_{s0} \to D_s \pi^0$ is enhanced with
respect to the radiative mode $D_{s0} \to D_s^* \gamma$.  The
estimated width of the meson is
 $\Gamma(D_{s0})\simeq 7$ KeV.
\end{quotation}
\newpage
\baselineskip=18pt
\vspace{2cm}
\section{Introduction}
The BaBar Collaboration has reported the observation of a narrow
peak in the $D_s^+ \pi^0$ invariant mass distribution,
corresponding to a state of mass $2.317$ GeV \cite{Aubert:2003fg}.
The  observed width is consistent with the resolution of the
detector, thus $\Gamma\le 10$ MeV.
In the same  analysis no  significant signals are found in
the $D_s \gamma$ and  $D_s \gamma \gamma$ mass distributions. The
meson has been denoted as $D_{sJ}(2317)$; the announcement
has immediately prompted different interpretations
\cite{many,eichten}.

A possible quantum number assignment to $D_{sJ}(2317)$ is
$J^P=0^+$, as suggested by the angular distribution of the meson
decay with respect to its direction in the $e^+-e^-$ center of
mass frame. This assignment can identify the meson with the
$D_{s0}$ state in the spectrum of the $c \bar s$ system.
Considering the masses of the other observed states belonging to
the same system, $D_{s1}(2536)$ and $D_{sJ}(2573)$, the mass of
the scalar $D_{s0}$ meson was expected in the range $2.45-2.5$
GeV, therefore $\sim 150$ MeV higher than the observed $2.317$
GeV. A $D_{s0}$ meson with such a large mass would be above the
threshold $M_{DK}=2.359$ GeV to strongly decay by $S$-wave Kaon
emission to $DK$, with a consequent broad width. For a  mass below
the $DK$ threshold the meson has to decay by different modes,
namely the isospin-breaking $D_s \pi^0$ mode observed by BaBar, or
radiatively. The $J^P=0^+$ assignment excludes the final state
$D_s \gamma$, due to angular momentum and parity conservation;
indeed such a final state has not been observed. On the other
hand, for a scalar $c \bar s$ meson the decay $D_{s0} \to D^*_s
\gamma$ is allowed. However, no evidence is reported yet of the
$D_s \gamma \gamma$ final state resulting from the decay chain
$D_{s0} \to D^*_s \gamma \to D_s \gamma \gamma$. In order to
confirm the identification of $D_{sJ}(2317)$ with the scalar
$D_{s0}$, one has at first to understand whether the decay modes
of a scalar particle with mass of $2317$ GeV can be predicted in
agreement with the experimental findings presently available. In
particular, the isospin violating decay to $D_s \pi^0$ should
proceed at a rate larger than the radiative mode $D_{s0} \to D_s^*
\gamma$, though not exceeding the experimental upper bound $\Gamma
\le 10$ MeV. This letter is devoted to such an issue.

\section{Mode $D_{s0} \to D_s \pi^0$}

In order to analyze the isospin violating transition
$D_{s0} \to D_s \pi^0$ one can use a formalism that  accounts for
the heavy quark spin-flavour symmetries in hadrons
containing a single heavy quark, and the chiral symmetry in the interaction
with the octet of light pseudoscalar states.

In the heavy quark limit, the heavy quark spin $\vec s_Q$ and the
light degrees of freedom total angular momentum $\vec s_\ell$ are
separately conserved. This allows to classify hadrons with a
single heavy quark $Q$ in terms of  $s_\ell$ by collecting them in
doublets the members of which only differ for the relative
orientation of $\vec s_Q$ and $\vec s_\ell$.

The doublets with $J^P=(0^-,1^-)$ and  $J^P=(0^+,1^+)$
(corresponding to $s_\ell^P= {1\over 2}^-$ and $s_\ell^P= {1\over
2}^+$, respectively) can be described by the effective fields
\begin{eqnarray}
H_a &=&
\frac{(1+{\rlap{v}/})}{2}[P_{a\mu}^*\gamma^\mu-P_a\gamma_5]
\label{neg} \\
S_a&=&\frac{1+{\rlap{v}/}}{2}
\left[P_{1a}^{\prime\mu}\gamma_\mu\gamma_5-P_{0a}\right]
\label{pos}
\end{eqnarray}
where $v$ is the four-velocity of the meson and $a$ is a light quark
flavour index. In particular in the charm sector the components of
the field  $H_a$  are $P_a^{(*)}=D^{(*)0},D^{(*)+}$ and
$D^{(*)}_s$  (for $a=1,2,3$); analogously, the components of $S_a$
are $P_{0a}=D^{0}_0,D^{+}_0,D_{s0}$ and $P_{1a}^{\prime}=D^{\prime
0}_1,D^{\prime +}_1,D^\prime_{s1}$.

In terms of these fields it is possible to build up an effective
Lagrange density describing the low energy interactions of heavy
mesons  with the pseudo Goldstone $\pi$, $K$ and $\eta$ bosons
\cite{hqet_chir,Casalbuoni:1992gi,Colangelo:1995ph,Casalbuoni:1996pg}:
\begin{eqnarray}
{\cal L} &=&
i\; Tr\{ H_b v^\mu D_{\mu ba} {\overline H}_a \}
+ \frac{f_\pi^2}{8} Tr\{\partial^\mu\Sigma\partial_\mu \Sigma^\dagger \}
+  Tr\{ S_b \;( i \; v^\mu D_{\mu ba} \; - \; \delta_{ba} \; \Delta)
{\overline S}_a \} \nonumber\\
&+ & \; i \;
g \; Tr\{H_b \gamma_\mu \gamma_5 {\cal A}^\mu_{ba} {\overline H}_a\}
 +  i \; g' \; Tr\{S_b \gamma_\mu \gamma_5 {\cal A}^\mu_{ba} {\overline S}_a\}
\nonumber \\
&+&\,[ i \, h \; Tr\{S_b \gamma_\mu \gamma_5 {\cal A}^\mu_{ba} {\overline H}_a\}
 \;  + \; h.c.] \;\;\; .  \label{L}
\end{eqnarray}
In  (\ref{L}) $\overline H_a$ and  $\overline S_a$ are defined as
$\overline H_a= \gamma^0 H^\dagger_a \gamma^0$ and $\overline S_a=
\gamma^0 S^\dagger_a \gamma^0$;   all the heavy field operators
contain a factor $\sqrt{M_P}$ and have dimension $3/2$. The
parameter $\Delta$ represents the mass splitting between positive
and negative parity states.

The $\pi$, $K$ and $\eta$ pseudo Goldstone bosons are included in
the effective lagrangian (\ref{L}) through the field
$\displaystyle \xi=e^{i {\cal M} \over f}$ that represents a
unitary matrix describing the pseudoscalar octet, with
\begin{equation}
{\cal M}=
\left (\begin{array}{ccc}
\sqrt{\frac{1}{2}}\pi^0+\sqrt{\frac{1}{6}}\eta & \pi^+ & K^+\nonumber\\
\pi^- & -\sqrt{\frac{1}{2}}\pi^0+\sqrt{\frac{1}{6}}\eta & K^0\\
K^- & {\bar K}^0 &-\sqrt{\frac{2}{3}}\eta
\end{array}\right ) \label{M}
\end{equation}
and $f\simeq f_{\pi}$. In eq.(\ref{L}) $\Sigma=\xi^2$,
while the operators $D$ and $\cal A$  are given by:
\begin{eqnarray}
D_{\mu ba}&=&\delta_{ba}\partial_\mu+{\cal V}_{\mu ba}
=\delta_{ba}\partial_\mu+\frac{1}{2}\left(\xi^\dagger\partial_\mu \xi
+\xi\partial_\mu \xi^\dagger\right)_{ba}\\
{\cal A}_{\mu ba}&=&\frac{1}{2}\left(\xi^\dagger\partial_\mu \xi-\xi
\partial_\mu \xi^\dagger\right)_{ba} \; .
\end{eqnarray}

The strong interactions between the heavy $H_a$ and $S_a$ mesons
with the light pseudoscalar mesons are
thus governed, in the heavy quark limit,  by three dimensionless
couplings: $g$, $h$ and $g^\prime$. In particular, $h$ describes
the coupling between a member of the $H_a$ doublet and one of the
$S_a$ doublet to a light pseudoscalar meson, and is the one
relevant for our discussion.

Isospin violation enters in the low energy Lagrangian of $\pi$,
K and $\eta$ mesons through the mass term
\be
{\cal L}_{mass} = {\tilde \mu f^2 \over 4}
Tr\{\xi m_q \xi +\xi^\dagger m_q \xi^\dagger\}  \label{Lmass}
\ee
with $m_q$ the light quark mass matrix:
\begin{equation}
m_q= \left (\begin{array}{ccc}
m_u&0&0\nonumber\\0&m_d&0\\ 0&0&m_s
\end{array}\right ) \,\,\, .
\end{equation}
In addition to the light meson mass terms, ${\cal L}_{mass}$
contains an interaction term between $\pi^0$ ($I=1$) and  $\eta$
($I=0$) mesons: ${\cal L}_{mixing}={\tilde \mu \over 2}
{m_d-m_u\over \sqrt 3} \pi^0 \eta$  which vanishes in the limit
$m_u=m_d$. As in the case of $D_s^* \to D_s \pi^0$ studied in
\cite{Cho:1994zu}, the isospin mixing term can drive the $D_{s0}
\to D_s \pi^0$ transition.\footnote{Electromagnetic contributions
to $D_{s0} \to D_s \pi^0$ are expected to be suppressed with
respect to the strong interaction mechanism considered here.} The
amplitude $A(D_{s0}\to D_s \pi^0)$ is simply written in terms of
$A(D_{s0}\to D_s \eta)$ obtained from (\ref{L}), $A(\eta\to
\pi^0)$ from (\ref{Lmass}) and the $\eta$ propagator that puts the
strange quark mass in the game. The resulting expression for the
decay amplitude involves the coupling $h$ and the suppression
factor $(m_d-m_u)/(m_s-{m_d+m_u \over 2})$ accounting for isospin
violation, so that the width $\Gamma(D_{s0} \to D_s \pi^0)$ reads:
\be \Gamma(D_{s0} \to D_s \pi^0) = {1 \over 16 \pi} {h^2 \over
f^2} {M_{D_s} \over M_{D_{s0}}} \Big( {m_d-m_u \over m_s-{m_d+m_u
\over 2}}\Big)^2 (1 + {m_{\pi^0}^2 \over |\vec p_{\pi^0}|^2})
|\vec p_{\pi^0}|^3 \,\,\, .\label{gamma1} \ee As for $h$, the
result of  QCD sum rule analyses of various heavy-light quark
current correlators is $|h| = 0.6 \pm 0.2$
\cite{Colangelo:1995ph}. Using the central value, together with
the factor $(m_d-m_u)/(m_s-{m_d+m_u \over 2}) \simeq {1 \over
43.7}$ \cite{Gasser:1984gg} and $f=f_\pi=132$ MeV we obtain: \be
\Gamma(D_{s0} \to D_s \pi^0)\simeq 6 \,\, KeV \label{had-ris} \,.
\ee Eq.(\ref{gamma1}) can receive $SU(3)_F$ corrections: a hint on
their size comes from the use of $f=f_\eta=171 \; MeV$ instead of
$f_\pi$ in (\ref{gamma1}), which gives $\Gamma(D_{s0} \to D_s
\pi^0)\simeq 4 \,\, KeV$. On the other hand, we neglect
corrections related to the finite charm quark mass.

The analogous calculation for $D_{s}^* \to D_s \pi^0$ involves the
coupling $g$ in (\ref{L}). Since $h$ and $g$ have similar sizes
($0.3\le g \le 0.6$), it turns out that the transitions $D_{s0}
\to D_s \pi^0$ is enhanced with respect to $D_{s}^* \to D_s \pi^0$
essentially due to kinematics, being ${|\vec p_{\pi^0}(D_{s0} \to
D_s \pi^0)|^3 \over |\vec p_{\pi^0}(D_{s}^* \to D_s \pi^0)|^3}
\simeq 3 \times 10^2$.

\section{Radiative $D_{s0} \to D_s^* \gamma $ decay}

Let us now turn to $D_{s0} \to D_s^* \gamma$, the amplitude of
which has the form:
  \be {\cal A}(D_{s0} \to D^*_s \gamma)= e \,
d \, [(\epsilon^* \cdot \eta^*)(p \cdot k)-(\eta^* \cdot p)
(\epsilon^* \cdot k)] \label{amptot} \,, \ee where $p$ is the
$D_{s0}$ momentum, $\epsilon$ the $D^*_s$ polarization vector, and
$k$ and $\eta$ the photon momentum and polarization. The
corresponding decay rate is: \be \Gamma(D_{s0} \to D_s^*
\gamma)=\alpha |d|^2 |\vec{k}|^3 \,\,\, . \label{radwid} \ee The
parameter $d$ gets contributions from the photon couplings to the
light quark part $e_s {\bar s}\gamma_\mu s$ and to the heavy quark
part $e_c {\bar c}\gamma_\mu c$ of the electromagnetic current,
$e_s$ and $e_c$ being strange and charm quark charges in units of
$e$. Its general structure is: \be d=d^{(h)}+d^{(\ell)}={e_c
\over \Lambda_c} + {e_s \over \Lambda_s} \, , \label{mu} \ee
where $\Lambda_a$ ($a=c,s$) have dimension of a mass. Such a
structure is already known from the constituent quark model. In the case
of $M1$ heavy meson transitions, an analogous structure
predicts a relative suppression of the radiative rate of the charged
$D^*$ mesons  with respect to the neutral one
\cite{Eichten:1979ms,Amundson:1992yp,Colangelo:1994jc,Colangelo:1993zq},
suppression that has  been experimentally confirmed
\cite{Hagiwara:fs}. From  (\ref{radwid},\ref{mu}) one could expect
a small width for the transition $D_{s0} \to D_s^* \gamma$, to be
compared to the hadronic width $D_{s0} \to D_s \pi^0$ which is
suppressed as well.

In order to determine the amplitude of  $D_{s0} \to D_s^* \gamma$
we follow a method based again on the use of heavy quark
symmetries, together with the vector meson dominance (VMD)
ansatz \cite{Amundson:1992yp,Colangelo:1993zq}. We first consider the coupling
of the photon to the heavy quark part of the e.m. current. The matrix element
$
\langle D^*_s(v^\prime, \epsilon)|{\bar c}\gamma_\mu c|D_{s0}(v) \rangle $
($v$, $v^\prime$ meson four-velocities)
can be computed in the heavy quark limit, matching the QCD
${\bar c}\gamma_\mu c$ current
onto the corresponding HQET expression \cite{Falk:1990pz}:
\be J_\mu^{HQET}=\bar h_v [ v_\mu +
{i \over 2 m_Q}( \overrightarrow \partial_\mu - \overleftarrow \partial_\mu) +
{i \over 2 m_Q} \sigma_{\mu\nu} ( \overrightarrow \partial^\nu + \overleftarrow \partial^\nu) + \dots]h_v \ee
where $h_v$ is the effective field of the heavy quark.
For transitions involving $D_{s0}$ and $D_s^*$, and for
$v=v^\prime$ ($v \cdot v^\prime=1$), 
the matrix element of $J_\mu^{HQET}$ vanishes. The consequence
is that $d^{(h)}$ is proportional to the inverse heavy quark mass
$m_Q$ and presents a suppression factor since in the physical
case
$v\cdot v^\prime=(m_{D_{s0}}^2+m_{D^*_s}^2)/2 m_{D_{s0}}m_{D^*_s}=1.004$.
Therefore, we neglect $d^{(h)}$ in (\ref{mu}).

To evaluate the coupling of the photon to the light quark part of
the electromagnetic current we invoke the VMD
ansatz and consider the contribution of the intermediate $\phi(1020)$:
\be \langle D^*_s(v^\prime, \epsilon)|{\bar
s}\gamma_\mu s|D_{s0}(v) \rangle=\sum_{\lambda}
 \langle D^*_s(v^\prime,
\epsilon) \phi(k, \epsilon_1(\lambda))|D_{s0}(v) \rangle {i \over k^2-
M_\phi^2} \langle 0 |{\bar s}\gamma_\mu s|\phi(k, \epsilon_1(\lambda))
\rangle \label{vmd} \ee with $k^2=0$ and $\langle 0 |{\bar
s}\gamma_\mu s|\phi(k, \epsilon_1) \rangle=M_\phi f_\phi
\epsilon_{1 \mu}$. The experimental value of $f_\phi$ is
$f_\phi=234 \, MeV$. The matrix element $\langle
D^*_s(v^\prime, \epsilon) \phi(k, \epsilon_1)|D_{s0}(v) \rangle$
describes the strong interaction of a light vector meson ($\phi$) with
two heavy mesons ($D^*_s$ and $D_{s0}$). It  can also be obtained through
a low energy  lagrangian in which the heavy fields $H_a$ and $S_a$ are
coupled, this time, to the octet of light vector mesons.\footnote{The
standard $\omega_8-\omega_0$ mixing is assumed, resulting
in a pure $\bar s s$ structure for $\phi$.}
The  Lagrange density is set up using
the hidden gauge symmetry method \cite{Casalbuoni:1992gi}, with
the light vector mesons collected in a 3 $\times$ 3  matrix
${\hat \rho}_\mu$ analogous to ${\cal M}$ in (\ref{M}).
The lagrangian\footnote{The role of other possible structures in
the effective lagrangian contributing to radiative decays is
discussed in \cite{Colangelo:1993zq}.} reads as \cite{Casalbuoni:1992dx}:
\be {\cal L}^\prime= i \,
\hat \mu \, Tr \{ {\bar S}_a H_b \sigma^{\lambda \nu} V_{\lambda
\nu}(\rho)_{ba} \} +h.c. \label{vec-lag} \,, \ee with $V_{\lambda
\nu}(\rho)=\partial_\lambda \rho_\nu-\partial_\nu
\rho_\lambda+[\rho_\lambda,\rho_\nu]$ and  $\rho_\lambda=i{g_V \over
\sqrt{2}}{\hat \rho}_\lambda$,
$g_V$ being fixed  to
$g_V=5.8$ by the KSRF rule \cite{Ksrf}. The coupling $\hat \mu$
in (\ref{vec-lag}) is constrained to $\hat \mu=-0.13 \pm 0.05
\, GeV^{-1}$ by the analysis of the $D \to K^*$ semileptonic
transitions induced by the axial weak current
\cite{Casalbuoni:1992dx}. The resulting expression for ${1 \over {\Lambda}_s}$
is:
\be {1 \over {\Lambda}_s}=-4\hat\mu{g_V \over
 \sqrt{2}} \sqrt{{M_{D^*_s}\over M_{D_{s0}}}}{f_\phi \over M_\phi}
\,\,\, .\label{mag-mom}
\ee
The parameters are obtained from independent channels;
we use their central values.

The numerical result for the radiative width:
\be \Gamma(D_{s0} \to D_s^* \gamma)
\simeq 1 \,\, KeV \label{rad-ris} \,  \ee
shows that the
hadronic $D_{s0} \to D_s^* \pi^0$ transition is  more
probable than the radiative mode $D_{s0} \to D_s^* \gamma$. In
particular, if we assume that the two modes essentially saturate
the  $D_{s0}$ width, we have
\be \Gamma(D_{s0})\simeq 7 \,\,\ KeV \ee \label{results}
 and \bea
{\cal B}(D_{s0} \to D_s \pi^0)&\simeq&0.85 \noindent \label{ratios}\\
{\cal B}(D_{s0} \to D_s^* \gamma)&\simeq&0.15  \nonumber \eea at odds with
the case of the $D_s^*$ meson, where the radiative mode dominates
the decay rate.

The same conclusion concerning the hierarchy of $D_{s0} \to
D_s\pi^0$ versus $D_{s0} \to D_s^* \gamma$ is reached in
\cite{eichten} using the quark model. Since our calculation
is based on a different method,
the  $s_\ell^P= {1\over 2}^-$ and $s_\ell^P= {1\over 2}^+$
doublets being treated as uncorrelated multiplets, we find the agreement
noticeable.

\section{Conclusions and perspectives}

We have found that the observed narrow width and the enhancement
of the $D_s \pi^0$ decay mode are compatible with the
identification of $D_{sJ}(2317)$ with the scalar state belonging
to the $s_\ell^P={1 \over 2}^+$ doublet of the $c \bar s$
spectrum. However, this conclusion does not avoid other questions
raised by the BaBar observation, one being the low mass of the
state. We believe that such a particular issue requires additional
investigations. A second point is that the radiative mode,
although suppressed, is not negligible, and should be observed at a
level typically represented by the ratios in (\ref{ratios}).

The quantum number assignment has two main and rather straightforward
consequences. The first one is the existence of the axial vector
partner $D_{s1}^\prime$ belonging to the same spin doublet
$s_\ell^P={1\over2}^+$. Even in the case where the
hyperfine splittings between  positive and negative parity
states are similar: $M_{D_{s1}^\prime}-M_{D_{s0}}\simeq
M_{D^*_s}-M_{D_s}$,  this meson is below the $D^* K$ threshold.
Therefore, its hadronic decay to $D_s^* \pi^0$, at the rate \bea
\Gamma(D_{s1}^\prime \to D_s^* \pi^0) &=& {h^2 \over 48 \pi f^2}
{M_{D^*_s} \over M_{D_{s1}^\prime}} \Big({m_d-m_u \over m_s-{m_d+m_u
\over 2}}\Big)^2 [2 + {\big( M^2_{D^*_s} + M^2_{D_{s1}^\prime} -
M^2_{\pi^0} \Big)^2 \over
4 M^2_{D^*_s} M_{D_{s1}^\prime}}] \nonumber \\
&\times&(1 + {m_{\pi^0}^2 \over |\vec p_{\pi^0}|^2}) |\vec
p_{\pi^0}|^3 \simeq \Gamma(D_{s0} \to D_s \pi^0) \label{gamma2}
\eea would produce a narrow peak in the $D_s^* \pi^0$ mass
distribution. The confirmation of such a state, the existence of
which is suggested by the analysis of the $D_s \gamma \pi^0$ mass
distribution \cite{Aubert:2003fg}, will support the
interpretation.

The second consequence concerns the doublet of scalar and axial
vector mesons in the $b \bar s$ spectrum. Since the mass splitting
between $B$ and $D$ states is similar to the corresponding mass splitting
between $B_s$ and $D_s$ states, such mesons should be below the
$BK$ and $B^*K$ thresholds, thus producing narrow peaks in $B_s
\pi^0$ and $B_s^* \pi^0$ mass distributions,  with  rates resulting
from expressions analogous to (\ref{gamma1})-(\ref{gamma2}).

\vspace*{1cm}
\noindent {\bf Note added.}
When this work was completed, the CLEO Collaboration announced the
observation of a narrow resonance with mass 2.46 GeV in the $D_s^{*+}\pi^0$
final state and the confirmation of $D_{sJ}(2317)$ \cite{CLEO}.
Moreover, a  theoretical analysis based on the quark model
was posted on the Los Alamos arXive, with the same conclusions presented here
\cite{godfrey}.

\vspace*{1cm}
\noindent {\bf Acnowledgments.}
We acknowledge partial support from the EC Contract No.
HPRN-CT-2002-00311 (EURIDICE).

\end{document}